\documentclass[%
 reprint,
 amsmath,amssymb,
 aps,
 pra,
]{revtex4-1}

\usepackage{graphicx}
\usepackage{dcolumn}
\usepackage{bm}
\usepackage{balance}
\usepackage{latexsym}
\usepackage{wasysym}
\usepackage{color}

\def\crta{\vrule height1.41ex depth-1.27ex width0.34em}
\def\dj{d\kern-0.36em\crta}
\def\Crta{\vrule height1ex depth-0.86ex width0.4em}
\def\Dj{D\kern-0.73em\Crta\kern0.33em}
\begin{document}

\title{Automated generation of Kochen-Specker sets}

\author{Mladen Pavi{\v c}i{\'c}}
\email{mpavicic@irb.hr}
\homepage{http://www.irb.hr/users/mpavicic}
\affiliation{Department of Physics---Nanooptics, Faculty of Math. and Natural Sci.~I, Humboldt University of Berlin, Germany}
\affiliation{Center of Excellence CEMS, Photonics and Quantum Optics
  Unit, Ru\dj er Bo\v skovi\'c Institute, Zagreb, Croatia.}
\author{Mordecai Waegell}
\affiliation{Institute for Quantum Studies, Chapman University,
  Orange, CA 92866, U.S.A.}
\author{Norman D. Megill}
\affiliation{Boston Information Group, Lexington, MA 02420, U.S.A.}
\author{P.K. Aravind}
\affiliation{Physics Department, Worcester Polytechnic Institute,
  Worcester, MA 01609, U.S.A.}

\date{May 1, 2019}


\begin{abstract}
Quantum contextuality turns out to be a necessary resource for
universal quantum computation and also has applications in
quantum communication. Thus it becomes important to generate
contextual sets of arbitrary structure and complexity to enable a
variety of implementations. In recent years, such generation has
been done for contextual sets known as Kochen-Specker sets.
Up to now, two approaches have been used for massive generation
of non-isomorphic Kochen-Specker sets: exhaustive generation up
to a given size and downward generation from master sets and their
associated coordinatizations. Master sets were obtained earlier
from serendipitous or intuitive connections with polytopes or Pauli
operators, and more recently from arbitrary vector components using
an algorithm that generates orthogonal vector groupings from them.
However, both upward and downward generation face an inherent
exponential complexity barrier. In contrast, in this paper we present
methods and algorithms that we apply to downward generation that can
overcome the exponential barrier in many cases of interest. These
involve tailoring and manipulating Kochen-Specker master sets
obtained from a small number of simple vector components, filtered
by the features of the sets we aim to obtain. Some of the classes of
Kochen-Specker sets we generate contain all previously known ones,
and others are completely novel. We provide examples of both kinds
in 4- and 6-dim Hilbert spaces. We also give a brief introduction
for a wider audience and a novice reader.
\end{abstract}

\maketitle

\section{\label{sec:intro}Introduction}

Quantum contextuality arguably plays an important role in the
field of quantum communication and quantum computation. In this
paper we shall focus on the most explored and used contextuality
configurations---the so-called Kochen-Specker (KS) sets.
We begin with a review of some important concepts.

Imagine a system of nonlinear equations describing a relation
between vectors or states. Let the vectors be 3-dim (3D) ones,
forming triplets:
${\mathbf a}_l=a_{lx}{\mathbf i}+a_{ly}{\mathbf j}+a_{lz}{\mathbf k}$,
$l=1,2,3$, and let the relation be the orthogonality
${\mathbf a}_l\cdot{\mathbf a}_m=0$, $l,m=1,2,3$, $l\ne m$.
Let us have many such triplets interconnected within a system,
and let us ask ourselves whether we can assign 0 and 1 to them
in such a way that just one vector in any triplet is assigned 1.
Now we have two problems for a chosen system. One is the
assignment problem of finding a 0-1 distribution (or proving
its impossibility), and the other is the coordinatization problem
of finding vector components. Both are of exponential complexity.
All the supercomputers on Earth would take an ``age of the universe''
amount of CPU time to solve either of the two problems, even for
some of the simplest systems. Now imagine three dots (we call them
{\em vertices\/}) on a paper or a screen connected by a curve (we
call it an {\em edge\/}). Connect them so as to follow the pattern in
which the aforementioned vector triplets are connected within their
system. They will form a so-called {\em hypergraph\/} and it has
been recognized that the two representations are equivalent:
vertices correspond to vectors, and edges to orthogonalities.
And finding solutions of the assignment problem for
hypergraphs statistically proves to be a feasible task on
supercomputers and clusters, typically reducing 
``age of the universe'' run times to anywhere from CPU seconds to
a few CPU weeks depending on the size and nature of the hypergraphs
of interest. A hypergraph with a solution is a classical noncontextual
system, and the one without it is a quantum contextual KS set. The
coordinatization problem was tougher. Finding vector components that
can be assigned to vertices was still a task of exponential
complexity. A number of workarounds have been designed, until most
recently we turned the problem upside-down. Instead of searching for
vector components we might assign to chosen hypergraphs, we search
for hypergraphs we might assign to chosen vector components, the
so-called {\em master sets\/}. But even that does not enable a
satisfactory efficent generation of KS sets and to achieve this goal,
in the present paper, we elaborate on features, algorithms,
and methods which not only speed up the search for KS sets almost
exponentially but also enable (reasonably) arbitrary exhaustive
generation of KS sets from master sets. The details and references
are given below.

A series of experimental implementations of 4D KS sets have been
carried out recently, using photons
\cite{simon-zeil00,michler-zeil-00,amselem-cabello-09,liu-09,d-ambrosio-cabello-13,ks-exp-03},
neutrons
\cite{h-rauch06,cabello-fillip-rauch-08,b-rauch-09},
trapped ions \cite{k-cabello-blatt-09}, and
solid state molecular nuclear spins  \cite{moussa-09}. Sets in 6D
have been implemented via six paths \cite{lisonek-14,canas-cabello-14}
and in 8D using photons \cite{canas-cabello-8d-14}.

These experiments have been implemented either under
assumption of predetermined noncontextual values or under
the {\em faithful measurement condition}
\cite{barrett-kent-04,appleby-05} meaning
that they did not overcome the noise-induced finite-precision
loophole. It is, however, possible to formalize contextuality in a
noise-robust manner and develop experimental tests for them
\cite{spekkens-05,kunjwal-spekkens-15,mazurek,kunjwal-18-arxiv}.
It is also possible to modify the Kochen-Specker notion by
appealing to {\em ontological faithfulness} \cite{winter}.

Our theoretical interest in contextual KS sets is justified by the
following recent achievements. 

A connection between contextuality and universal quantum computation
has been established to an extent that makes contextuality a
necessary resource for universal quantum computation. A pressing
open question for qubits is whether a suitable operationally
motivated refinement or quantification of contextuality can provide
a quantum speed-up \cite{magic-14}. Also, one can design quantum
gates via operator-generated 4D complex KS sets
\cite{waeg-aravind-jpa-11}.

In quantum communication, KS sets might protect
\cite{cabello-dambrosio-11} and secure
\cite{nagata-05} quantum key distribution (QKD) protocols.

In lattice theory, KS sets have served as generators of
higher-order generalized orthoarguesian lattices
\cite{bdm-ndm-mp-fresl-jmp-10,mp-7oa}.

For the above applications of KS sets, especially in quantum
computation, it is important to generate sufficiently
large sets to enable varieties of different implementations
and to characterize their main features and information on their
structure.

To date, all papers on KS sets in the literature have been focused on

(i) discovering or implementing smallest sets (e.g.,
  \cite{cabell-est-96a}), or on

(ii)  an exhaustive upward hypergraph-generation of sets, e.g.,
\cite{pmmm05a}, or on

(iii)  a top-down, random downward generation of sets from
fortuitously obtained master sets (e.g., %
\cite{aravind10,waeg-aravind-jpa-11,mfwap-11,mp-nm-pka-mw-11,waeg-aravind-megill-pavicic-11,waegell-aravind-12,waeg-aravind-pra-13,waeg-aravind-fp-14,waeg-aravind-jpa-15,waeg-aravind-pla-17,pavicic-pra-17}), or on

(iv) generating sets in higher dimensions from the ones in
smaller dimensions (e.g.,
\cite{zimba-penrose,cabell-garc96,cabell-est-05}), or on

(v) generating big master sets from simple vector components
in order to compare them with master sets obtained via
(i-iii) methods (\cite{pm-entropy18}).

Advantages and disadvantages of these approaches are as follows.

(i) The smallest KS sets are of just historical relevance, because
  practically all of them in even dimensional (4D through 32D) Hilbert
  spaces are already known. Besides, all of them are but by-products
  of our current generation algorithms;

(ii) At least for the time being, exhaustive upward
  hypergraph-generation of KS sets faces computational limits of
  supercomputers and is limited to ca.~40 hypergraph vertices in 3D,
  to ca.~25 ones in 4D, and so on. Still, this kind of a generation
  remains the only deterministic and exhaustive one.

(iii) Big master sets which serve for random downward generation of
  smaller sets are based on serendipitous or intuitively found
  connections of KS hypergraphs with polytopes or Pauli operators
  which we run out of and which do not generalize.

(iv) No universal constructive algorithm for arbitrary input
  KS sets from smaller dimensions has been proposed for a
  potential automated generation of sets in higher dimensions.
  However, the approach might be best option for big sets and
  dimensions where other methods face computational limits.

(v) The algorithms and programs are computationally feasible for
  generating master sets from the vector components that are already
  known to yield KS masters obtained via other methods (i-iii), even
  on a PC (this is what has been carried out in
  Ref.~\cite{pm-entropy18}). However, they are not computationally
  feasible, even on supercomputers, for random computer generated
  components without additional analyses, algorithms, and
  programs we provide in the present paper. For instance, in
  Ref.~\cite{pm-entropy18}(Table 2) we were not able to obtain
  smaller KS sets from the 4D 2316-3052, or 6D 11808-314446,
  or 8D 3280-1361376, or 16D, or 32D master sets directly. The main
  reason is that we cannot generate smaller sets from a master set
  by a brute force---the programs are too slow for that.
  E.g., in 6D, without a new algorithm for splitting the original
  KS master 834-1609 (obtained from the vector components
  in Ref.~\cite{pm-entropy18}) into two smaller sub-masters
  (as we do here), we would not have been able to arrive at the
  distribution given in Fig.~\ref{figure-5}
  and results in Fig.~\ref{figure-4}(e,f,g). Taken together, 
  the main aim of Ref.~\cite{pm-entropy18} was to classify KS sets 
  in even dimensions via vector-component generated master sets that
  would contain all known KS sets, while the main aim of the present
  paper is to provide a functional and feasible method of generating
  KS sets of chosen sizes and properties from such master sets.
  
In the present paper, we present an optimized method of generating
KS sets from tweaked KS master sets engineered via the method
outlined in Ref.~\cite{pm-entropy18}. In the latter method, we start
with random vector components and build $n$-tuples of mutually
orthogonal vectors which results a in KS masters with a specific
coordinatization. In the former method, the components are filtered
by features of KS sets obtained from coordinatizations of already
known KS sets as exemplified below (see figures in the main body of
the paper and in the Appendices). This provides us not
only with a uniform and general method of computationally feasible
KS set generation, but also with a larger scope and a bigger, more
thorough, picture of quantum contextuality than any of the previous
approaches. That comes at a price of engaging supercomputers to
arrive at results, but once obtained, they can be verified either
by hand or via a graphical representation as given by our figures.

Experimentally, the complexity of KS set implementation grows only
linearly with the complexity of their structure and, on the other
hand, the simplest KS sets often do not possess features that
bigger KS sets exhibit, such as the so-called $\delta$-feature
\cite{pavicic-pra-17,pm-entropy18} (see Fig.~\ref{figure-3}
and defined in Subsecs.~\ref{subsec:mmp} and ~\ref{subsec:vgen}
below), absence of coordinatizations with real vectors, and the
absence of parity proofs.

Since the proposed approach charts a new territory, we limit
ourselves to 4D space and an example from 6D space, but we
stress that the approach can be applied to any dimension. Also,
we answer several recently posed open questions.

In this paper, to achieve automated generation of KS sets we rely
on our hypergraph language and its algorithms and C programs first
given in \cite{pmmm05a} and extended here, analogous to the recently
developed algorithm and program {\textsc{Melvin}}
\cite{krenn-zeilinger-16} whose authors carried out an
``Automated Search for New Quantum Experiments.'' We share their
guiding principle: ``In contrast to human designers of experiments,
[{\em our computer programs\/}] do not follow intuitive reasoning
about the physical system and, therefore, lead to the utilisation
of many unfamiliar and unconventional techniques that are
challenging to understand'' \cite{krenn-zeilinger-16}.

\section{\label{sec:res}Results}

The main result of the paper is that via a rather simple algorithm
we can generate any class of KS sets from basic real or complex
vectors. Let us consider the following example. In the last 27
years several research groups (Kochen, Specker, Peres, Cabello,
Pavi\v ci\'c, Megill, McKay, Merlet, Aravind, Waegell, and
others---see Fig.~1 and Sec.~III in \cite{pavicic-pra-17}) obtained
1233 4D KS sets by investing a considerable amount of research work
and CPU-centuries of calculation on supercomputers which they
published in over 20 papers. In contrast, with the help of our
algorithm, by just putting the three components of their vectors
\{-1,0,1\} in its program, we obtain all those KS sets and
additional ones containing them, in seconds on a PC.

So what is a KS set? It is a set of $n$-tuples of mutually orthogonal
vectors that provides a constructive proof of the Kochen-Specker
theorem. This theorem states that it is impossible to ascribe
predetermined eigenvalues to all quantum observables, which we call
the contextuality of quantum mechanics. We represent the $n$-tuples
as edges on a hypergraph, with individual vectors assigned to the
vertices of the hypergraph.

{\em Kochen-Specker theorem} \cite{koch-speck,zimba-penrose} reads:
In ${\cal H}^n$, $n\ge 3$, there are sets of $n$-tuples of mutually
orthogonal vectors to which it is impossible to assign 1s and 0s
in such a way that (i) No two orthogonal vectors are both
assigned the value 1 and (ii) In any group of $n$ mutually orthogonal
vectors, not all of the vectors are assigned the value 0.

These sets are called {\em KS sets\/} and the vectors
{\em KS vectors\/}.

That means that the terms ``generation of hypergraphs'' and
``downward generation'' used in the text are tantamount to
the term ``generation of KS sets'' used in any of the papers
mentioned in points (i-v) in Sec.~\ref{sec:intro} and to the
term ``generation of KS sets from master sets'' used in (iii)
and (v), respectively.

Generating hypergraphs and checking them for the KS
property are exponentially complex tasks in general but for the
great majority of runs they turn out to run in a feasible amount
of time due to various heuristics and other techniques
incorporated into our algorithms. So, for instance, our programs
can generate and assign vectors to hypergraphs with over 100,000
vertices and edges. Verification of the KS property becomes very
demanding on supercomputers when the number of vectors/vertices
exceeds 1,000 and the number of dimensions exceeds 10, though.

We limit ourselves to critical non-isomorphic KS sets, where
{\em critical\/} means that they are minimal in the sense that
removing any $n$-tuple of mutually orthogonal vectors
(i.e.\ any hypergraph edge) turns a KS set
into a non-KS set. In other words, they represent non-redundant
blueprints for their implementation since bigger KS sets that
contain them only add orthogonalities that do not change the KS
property of the critical sets. Note that edges within a hypergraph
correspond to measurements necessary to show contextuality.

To deal with hypergraphs conveniently in our computer programs,
we make use of the McKay-Megill-Pavi\v ci\'c (MMP)
hypergraph language \cite{pmmm05a,pavicic-pra-17}.

\subsection{\label{subsec:mmp}MMP Hypergraph Language}

We describe KS sets by means of McKay-Megill-Pavi\v ci\'c (MMP)
hypergraphs, which are defined as hypergraphs in which edges
contain at least $n$ vertices (corresponding to orthogonal
$n$-tuples of vectors) and intersect each other in at most
$n-2$ vertices (corresponding to vectors themselves). MMP
hypergraphs are encoded by means of printable ASCII characters
\cite{pmmm05a}.
Vertices are denoted by one of the following characters:
{{\tt 1 2 \dots 9 A B \dots Z a b
\dots z ! " \#} {\$} \% \& ' ( ) * - / : ; \textless\ =
\textgreater\ ? @ [ {$\backslash$} ] \^{} \_ {`} {\{}
{\textbar} \} $\sim$} \cite{pmmm05a}. When all of
them are exhausted we reuse them prefixed by `+',
then again by `++', and so forth. An $n$-dim KS set with $k$ vectors
and $m$ $n$-tuples is represented by an MMP hypergraph with
$k$ vertices an $m$ edges which we denote as a $k-m$ set.
A KS set with a parity proof is one with an odd number of edges,
in which all vertices belong to an even number of those edges,
which makes the impossibility of a 0-1 assignment transparent.

A number of examples are given below. We generate, process, and
handle MMP hypergraphs by means of algorithms in the programs
{\textsc{Shortd, Mmpstrip, Mmpsubgraph, Vecfind, States01}}, and
others. See the Appendices for more details.

Note that the aforementioned $n-2$-intersection condition is the
only condition that restricts a general hypergraph to an MMP
hypergraph \cite{pmmm05a}. It also directly generates the
$\delta$-feature (two edges might share $n-2$ vertices) we
introduce in Subsec.~\ref{subsec:vgen} below.

\subsection{\label{subsec:coord}Coordinatization of
KS Sets}

As we mentioned at the beginning of this section, we can
generate the aforementioned KS sets from nothing but three vector
components \{-1,0,1\} provided to program {\textsc{Vecfind}}.

It works as follows.  We specify a dimension $n$ (4 in this
example), a set of vector component values \{$0,\pm 1$\}, and
the option {\tt{master}}. (In general, the vector component
values can be arithmetic expressions involving complex numbers,
including exponential and trigonometric functions.) The program
builds an internal list of all possible non-zero vectors
containing these components. (The vectors are not normalized in
general. If two vectors are proportional, one of them is discarded.)
From this list, it finds all possible mutually orthogonal
$n$-tuples of vectors. It then generates a hypergraph
(in the MMP notation) with edges corresponding to these mutually
orthogonal $n$-tuples.  We call this hypergraph a ``master set.''
Depending on the vector component values chosen, this master set may
or may not have the KS property. The program {\textsc{States01}}
tests for this property, and master sets without the KS property are
discarded.  When that happens, we can add more vector component
values or try a different set of vector component values. Once we
obtain a master set with the KS property, we can use our already
existing programs (developed for our previous work with master sets
obtained from polytopes, etc.) to extract critical KS subsets.
Although in one sense the {\tt{master}} option algorithm is
straightforward---it just exhausts all possible mutually orthogonal
$n$-tuples that can be built from given vector components---it
nonetheless creates master sets that contain all the master sets of
our previous work derived from polytopes or Pauli operators,
when we provide it with nothing but the same vector components.

In this example, the set we obtain is indeed a KS master set with
40 vertices and 32 edges (See Fig.~\ref{figure-6}).
It reveals that Peres' 24-24 set obtained previously is not a
{\em proper\/} master set in the sense of generating all KS subsets
with the coordinatization (assignment of vectors) generated by the
vector components of the Peres coordinatization. Details are given
in Appendix \ref{sec:app1}.

\subsection{\label{subsec:polyt}Polytope vs.~Vector
Generation}

In the last ten years, various KS sets have been generated from a
number of polytopes for which correspondences with KS hypergraphs
have been found
\cite{aravind10,waeg-aravind-jpa-11,mfwap-11,mp-nm-pka-mw-11,waeg-aravind-megill-pavicic-11,waegell-aravind-12,waeg-aravind-pra-13,waeg-aravind-fp-14,waeg-aravind-jpa-15,waeg-aravind-pla-17}.
These correspondences have proven very useful in the development
of the more systematic approach to the generation of KS sets
developed in this paper.

The vertices of regular or highly symmetrical polytopes in a number
of dimensions, in both real and complex spaces, can be mapped into
numerous KS master sets in Hilbert spaces. In particular, the 24-cell,
600-cell and 120-cell, which are the three exceptional 4D regular
polytopes (having no analogs in lower or higher dimensions) provide
striking illustrations of this statement. All these polytopes have
the feature that their vertices lie on the surfaces of 3D-spheres and
come in antipodal pairs whose members are diametrically opposite
each other on the 3D-sphere. Since each antipodal pair maps on to a
single ray, each of these polytopes gives rise to half as many rays
as its vertices. The 24-, 600- and 120-cells thus lead to 12, 60 and
300 rays, respectively. 

Furthermore, the vertices of each of these polytopes contain many
inscribed copies of the cross-polytope (or 16-cell), each of which
give rise to 4 mutually orthogonal rays, and thus corresponds to an
edge in an MMP diagram. Naturally, only polytopes which give rise
to edges can give rise to KS sets, and so, not all polytopes have
a direct mapping to a KS master, nor do all KS masters have a
recognized mapping from a polytope.

In the case of the 24-cell, one must unite the 12 rays with a second
set of 12 rays obtained from its dual (another 24-cell, but oriented
differently from the first) to obtain a set of 24 vertices and 24
edges that is identical to the famous 24-24 set in 4-dimensions
discovered originally by Peres and then explored by a number of other
authors.

The 600-cell gives rise to a 4D KS master set of 60 vertices that
form 75 edges of mutually orthogonal vertices. By stripping only one
edge at the time from this set using MMPSTRIP, we obtain 75
isomorphic 60-74 KS subsets which merge into a single 60-74 set.
Their subsets can therefore be called the 60-75 and 60-74
classes, respectively. They are identical, short of the 60-75 set
from the 60-75 class [38].

The 120-cell, which is the dual of the 600-cell, is unusual in that
it contains ten 600-cells as subsets within it. Further, these ten
600-cells divide into two sets of five, with the members of each set
having no vertices in common and the two together covering all the
vertices of the 120-cell. From this it follows that the
corresponding quantum 300-675 KS set has ten different 60-75 sets
within it as subsets.

It should be noted that the members of the 60-74(75) class have
vector components from the set
${\cal V}=\{0,\pm(\sqrt{5}-1)/2,\pm 1,\pm(\sqrt{5}+1)/2,2\}$
\cite{waeg-aravind-megill-pavicic-11},
while those of the 300-675 class have vector components derived
from the elements of matrices based on
$\cal V$
\cite{waeg-aravind-fp-14}.

This might suggest that the vector sets generated from the
aforementioned coordinates would lead to not just the 24-24,
60-74(75) and 300-675 classes, but also the inclusion of the
second of these within the third. However this turns out not
to be the case at all. The set $\cal V$ unexpectedly generates
a KS master set, 2316-3052, consisting of 2316 vertices and
3052 edges, whose class of KS sets is too large to be determined
by the present methods. But what is even more interesting is that
we need not employ all the components of $\cal V$ to get the
members of the 60-74 class. Removing just 2 from $\cal V$ gives
us a 888-1080 master set, the further removal of $(\sqrt{5}+1)/2$
gives a 676-848 master set, the still further removal of
$(-\sqrt{5}-1)/2$ $\to$ a 272-268 master set and, finally, the
removal of $(-\sqrt{5}+1)/2$ $\to$ a 156-120 master set. In
other words, we have the nesting of master sets described by the
relation 156-120 $\subset$ 272-268 $\subset$ 676-848 $\subset$
888-1080 $\subset$ 2316-3052, with the largest of these having
all elements of $\cal V$ as coordinates of its vectors and each
member of the sequence having one less element than the member
above it.

\begin{figure*}[htp]
\begin{center}
\includegraphics[width=.99\textwidth]{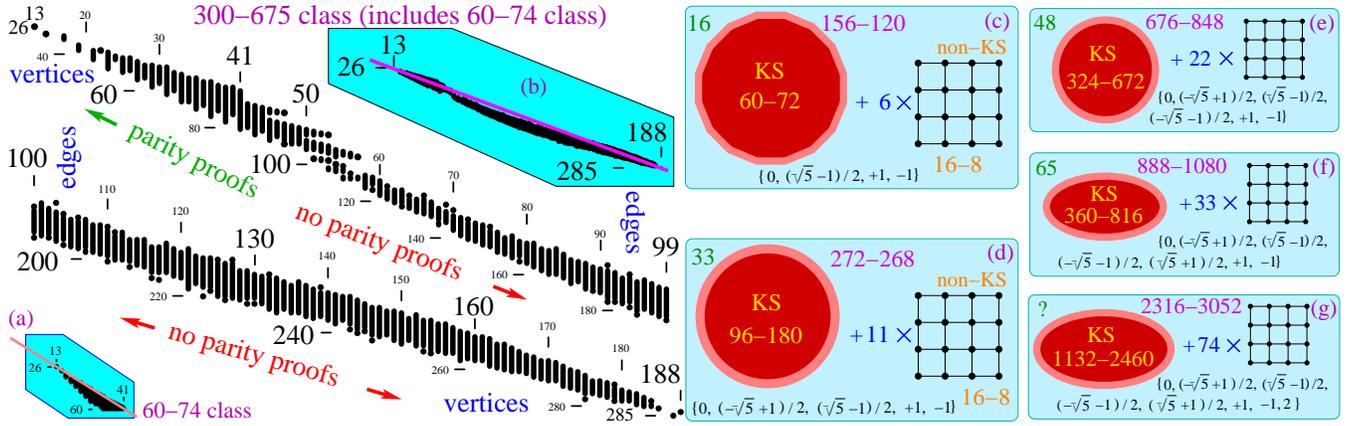}
\end{center}
\caption{Distribution of $10^{10}$ non-isomorphic critical KS sets
  from the 300-675 class we found by means of programs
  {\textsc{Mmpstrip, States01}}, and {\textsc{Mmpsubgraph}}.
  Dots represent KS criticals with the number of edges given
  at the ordinates and the number of vertices at the abscissae.
  When dots are next to each other (``glued'' together), they are
  shown as vertical lines; e.g., at the far right end of the
  upper strip, there is a line of KS criticals with 99 edges ranging
  from 168 to 183 vertices, then, just below it, there is a 184-99
  gap, followed by a dot which represent KS criticals with 99 edges
  and 185 vertices. Vertex units are scaled to 1/4 of edge units
  for a compact presentation. Inset (b) gives the whole span of
  criticals and shows that the criticals are agglomerated below the
  line which connects the smallest and largest criticals.
  Insets (c)-(g) show vector generated sets with the given 
  coordinatization: KS hypergraphs for each of them are represented
  via biggest outer polygon loops found by {\textsc{Loop}}
  (after searching through 50,000 combinations for each):
  16 edges (c), 33 (d), 48 (e), 65 (f), and `?' (g), indicating
  that we did not run {\textsc{Loop}} on them because it would
  take too much CPU time---cf.~hexagon 18-9 in Fig.~\ref{figure-6}.}
\label{figure-1}
\end{figure*}

As was the case with the 40-32 master set discussed in the
previous subsection, each of the five master sets just described
can be decomposed into two disconnected pieces. The smallest of
them, 156-120, consists of a 60-72 KS set and 6 structurally
identical non-KS 16-8 sets, which is conveniently expressed by
the equation $156-120=60-72 + 6\times 16-8$. In an exactly similar
way, the decompositions of the other master sets can be described
by the equations given in Fig.~\ref{figure-1}. All 16-8 sets have
the same hypergraph representation as the one shown in the right
part of Fig.~\ref{figure-6}, i.e., they are all isomorphic to
each other.

The above master sets might well contain the 60-75 and 300-675
classes, but we did not pursue this point (although we did verify
that they contained a large number of KS sets from these classes).
We demonstrated earlier that the 300-675 class contained both the
60-74(75) and 96-96 classes [35], and filled in the gap between
the 82-41 and 211-827 sets we were left with in [38]. This allowed
us to assemble all the information we have presented in
Fig.~\ref{figure-1} and in the Appendices.

In Fig.~\ref{figure-1} we see that the 60-74 class exhibits a
{\em cut-off\/} feature from below which we find, via both
the polytope and vector generation, to be an indication of the
existence of a bigger set which contains it. There are actually
several such cut-offs in the figure which form saw-tooths
at the levels of 60, 72, 84, and 96 vertices which might
correspond to the consecutive vector generated sets above.
There is also a spindle-shaped agglomeration of a more complete
distribution of sets as shown in Fig.~\ref{figure-1}. These
features are elaborated in detail in the Appendices,
and they are our starting points for a pure vector generation of
supersets in the next subsection.

\subsection{\label{subsec:vgen}Vector Generation of
  KS Sets from Basic Vector Components}

The theoretical background and algorithms of quantum computation
are to a large extent based on qubits and their states,
including entanglements in particular. KS sets
should therefore also allow a qubit state representation.
This can be done so as to define them via operators whose
eigenvectors (eigenstates) then directly define
vertices of MMP hypergraphs in the complex 4D
${\mathcal{H}}^2\otimes{\mathcal{H}}^2$ Hilbert space.
Via that approach we shall get the coordinatization for our
vector generation.

We carried out the qubit approach in \cite{waeg-aravind-jpa-11}
by means of Pauli operators for two qubits and their eigenstates
and generated the 60-105 KS class. In \cite{waeg-aravind-jpa-11}
we obtained KS sets with an odd number of edges with the help
of parity proof programs, and in \cite{pavicic-pra-17} we
generated sets with even numbers of edges and also
all those ones with an odd number of edges which do not have
parity proofs via our MMPSTRIP and STATES01 programs. The master
set in MMP hypergraph notation is given in
\cite[App.~2, p.~22]{pavicic-pra-17}. A number of 60-105
KS class figures and MMP hypergraphs are given in
\cite{pavicic-pra-17}.

The (eigen)vectors we obtained have components in the complex field,
while Peres' 24-24 set and the whole 24-24 class is properly contained
in the 60-105 class, which originally used real coordinatization.
Bigger sets from the 60-105 class do not allow simple real
coordinatization, e.g., with components from \{-1,0,1\} as we
have shown in the Appendices, but they might allow an
intricate kind of real coordinatization which are not eigenvectors
of the Pauli operators that we used to generate the 60-105 KS class.
So, such real vectors might allow a representation by means of
particles with four level spin $s=\frac{3}{2}$ (verifiable via,
e.g., a Stern-Gerlach device), so that $\dim{\mathcal{H}}_s=2s+1$
is satisfied, but possibly not via two qubits
($\dim({\mathcal{H}}^2\otimes{\mathcal{H}}^2)=2^2=4$).
This prompted us to investigate to what extent KS sets are
determined by coordinatizations that they do or do not allow, and
whether a coordinatization itself can enable us to generate
bigger sets and master sets.

\begin{figure*}[htp]
\includegraphics[width=\textwidth]{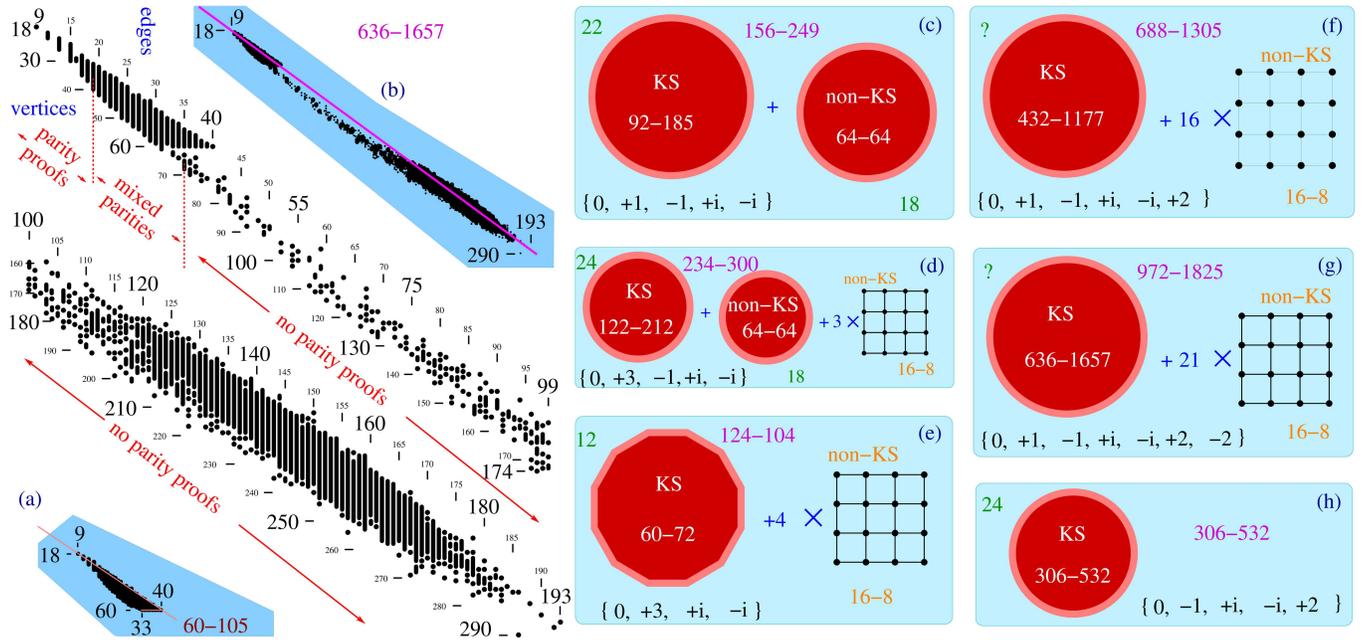}
\caption{Distribution of KS non-isomorphic criticals from the
  636-1657 class. 7,720,530 criticals from the 60-105 subclass
  are taken over from Fig.~4 of \cite{pavicic-pra-17} and
  12,952 higher criticals are generated for this paper.
  For the representation of the criticals via dots and lines
  in the main figure, see the caption of Fig.~\ref{figure-1}.
  The 60-105 KS criticals exhibit a flat cutoff feature
  at the level of 60 vertices, inset (a), similar to the one
  shown in the inset (a) of Fig.~\ref{figure-1}; The global
  distribution is shown in the inset (b); smaller master sets
  generated from smaller sets of vector components and mostly
  consisting of a KS set and several isomorphic non-KS sets are
  shown in insets (c-h); (f,g) `?' again means that it would take
  too much CPU time to run {\textsc{Loop}} on the sets.}
\label{figure-2}
\end{figure*}

In \cite{waeg-aravind-jpa-11,pavicic-pra-17} we presented in
detail how the operator structure and related eigenstates
(eigenvectors, vertices) were obtained from the 600-cell polytope
and how the whole 60-105 class was generated and which features
its KS sets possess. The eigenvectors had the components from
the set ${\cal I}=\{0,\pm 1,\pm i\}$. However, the class also
exhibits a flat cutoff at the 60 vertex level of critical KS sets
shown in Fig.~\ref{figure-2}(a), similar to the one of the
60-74 class, shown in Fig.~\ref{figure-1}(a). That indicated that
${\cal I}$ might generate a much bigger master set containing the
60-105 one, although, to our knowledge, no one has come forward
with such a set. We are now confirming the conjecture by means
of a direct vector generation from $\cal I$.

So, we started our vector generation with components from
${\cal I}$ and then pursued its variations via an
automated computer search. From $\cal I$ we arrive
at the 156-249 master set whose KS subset 92-185 is already much
bigger than the 60-105 that is contained in it. By stripping down its
edges via MMPSTRIP and filtering with STATES01, we form the 92-185 class.
By adding 2 to $\cal I$ ($\{0,\pm 1,\pm i,2\}$), we get 688-1305, and
by adding -2 to it ($\{0,\pm 1,\pm i,\pm 2\}$) we obtain the 972-1852
master and the class 636-1657 which contains all the others. They are
shown in Fig.~\ref{figure-2}(c,f,g). As the master sets from the
previous section, they also consist of smaller KS and non-KS sets.
However, the non-KS sets are not always 16-8 ones but also 64-64
ones as in Fig.~\ref{figure-2}(c,d). All 16-8 sets are
isomorphic to each other, thus effectively reducing all the masters
to two sets: one KS and one non-KS. Sometimes we do not
get a non-KS set at all, as e,g., in
Fig.~\ref{figure-2}(h) when all the vectors are exhausted in
building the KS set. We were unable to find correlations between
kinds of non-KS set and vector components indicated at the bottom of
Fig.~\ref{figure-2}(c-h).

Each set of components yield a different coordinatization for the
same KS sets and therefore give different possible experimental
implementations. For instance, 21-11 can be assigned vectors
generated by the set of components \{0,+3,$\pm i$\} or
\{0,$\pm 1$,$\pm i$\} or any other from Fig.~\ref{figure-2}(c-h).
In Appendix A of Ref.~\cite{pm-entropy18} we show how components of
vectors of a given coordinatization of a 21-11 KS set directly
generate quantum states via which we might implement the set.

In Fig.~\ref{figure-2}, the hypergraphs from the 636-1657
class are not so uniformly distributed as those from the 60-105
simply because a generation from such a big master requires orders
of magnitude more time. Still, an especially sparse population of
hypergraphs with 38 to 117 edges indicates an uneven probability
of getting individual KS critical sets from the master set,
similarly to the 300-675 probabilities (recall the gaps
we obtained in \cite{megill-pavicic-mipro-17,pavicic-pra-17}).

Five smaller hypergraphs from the class 636-1657 that are not
subgraphs of 60-105 are given in Fig.~\ref{figure-3}.
What is interesting about them is that the first four of them,
although obtained from the $\{0,\pm 1,\pm i,\pm 2\}$ based master
set, also have a $\{0,\pm 1,\pm i\}$ coordinatization.
For instance, {\bf 22-11$+$}
{\tt 1234,4567,789A,ABCD,DCEF,FEGH,HIJK,KLM1,68JL, 29GM,35BI.
  \{1=\{1,1,1,1\},2=\{1,1,-1,-1\},3=\{1,-1,\break i,-i\},4=\{1,-1,-i,i\},5=\{1,1,i,i\},6=\{1,i,1,-i\},\break 7=\{-1,i,1,i\},8=\{1,i,-1,i\},9=\{1,0,1,0\},A=\{0,1,0,\break -1\},B=\{i,0,1,0\},C=\{-i,i,1,i\},D=\{1,1,i,1\},E=\{i,\break -i,1,i\},F=\{i,i,1,-i\},G=\{0,1,0,1\},H=\{1,0,i,0\},I=\break \{0,i,0,1\},J=\{i,1,1,i\},K=\{1,i,-i,-1\},L=\{1,-i,i,\break -1\},M=\{1,-1,-1,1\}\}}\ 
\ (`$+$' in {\bf 22-11$+$} means that it is a 2nd {\bf 22-11} critical
in the class).

The fifth hypergraph does not have
such a coordinatization: {\bf 25-13$+$2} \
{\tt 1234,4567,789A,ABCD,CDEF,EFGH,\break GHIJ,JKLM,MNO1,29BO,
  345P,4PIK,68LN.} (`$+${\bf 2}' in {\bf 22-13$+$2}
means that it is a 3rd {\bf 22-13} critical in the class).

The {\bf 61-31$+$} critical, with more vertices than
any KS set from the 60-105 subclass, possesses a
$\{0,\pm 1,\pm i,2\}$ coordinatization.
{\bf 61-31$+$} \ {\tt 1234,5674,89A7,BCD6,EFGH,\break IJKL,MNKL,OPMN,QR23,STUV,WXIJ,OPGH,UVDA,STC9,\break YZWX,abcd,eQRF,fgcd,hijk,lmkg,nojf,pome,qr51,\break stuZ,vurY,wabE,xtB8,ywvq,yxsp,zynl,zyhi.}

  \parindent=20pt

The biggest critical with 31 edges from the 60-105 subclass has
58 vertices.

There are also KS hypergraphs with more than 60 vertices and
with a $\{0,\pm 1,\pm i\}$ coordinatization, e.g., those
obtained from the 156-249 master. The following 61-33 is one of
the smallest of them. {\bf 61-33$+$} \
{\tt 1234,1256,789A,7BCD,8BE5,F9GH,\break FAIJ,KLMN,KLOP,QROS,QTUV,WRXY,WTZ4,abTc,adIe,\break aTfg,achi,bdE6,jklm,jlni,XopN,Zq34,oYfr,kmhs,\break q4IJ,Gtuv,wxtHwyIe,xyCD,Mpns,zOrg,zOSP,uUVv.}

\parindent=20pt

The KS sets above have parity proofs. Actually, all KS criticals
from 18-9 through 35-19 do have them, while none from 66-35 through
284-193 has one. In the region from 36-19 through 65-35
({\em mixed parities\/}), KS criticals with odd number of edges
might or might not have them (KS sets with even number of edges
cannot have a parity proof by definition). Parities are indicated
in Fig.~\ref{figure-2}.

Most of the hypergraphs from the 636-1657 class exhibit the
$\delta$-feature we recognized in Sec.~V and Fig.~5 of
\cite{pavicic-pra-17} and indicated in Fig.~\ref{figure-3}---when
pairs of edges share two vertices i.e. intersect each other
twice at two vertices.

Global distribution of the hypergraphs shown in the inset (b) of
Fig.~\ref{figure-2} exhibits the agglomeration below the
line which connects the smallest and largest criticals. Also, the
number of vertices per edge decreases toward both smallest and
biggest vertices. Both features are consistent with those of
300-675 and 148-265 classes shown in Figs.~\ref{figure-1}(b)
and \ref{figure-9}(b).

\begin{widetext}
\begin{figure*}
\begin{center}
  \includegraphics[width=.99\textwidth]{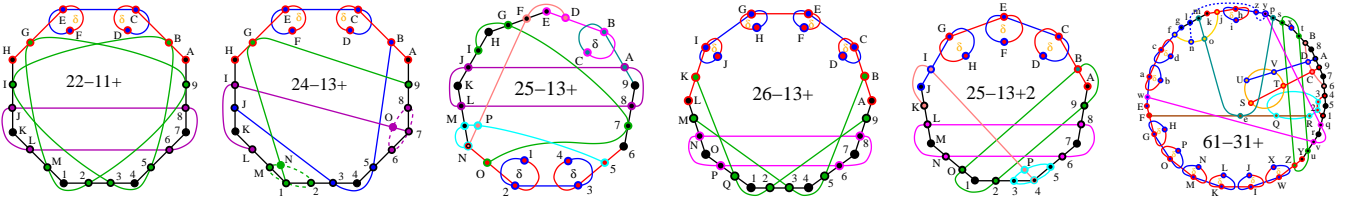}
\end{center}
\caption{MMP hypergraphs of KS critical sets derived from the
  636-1657 master that are not subgraphs of 60-105. The first
  four share the vertex components with the latter master while
  the last two do not; $\delta$'s denote the $\delta$-feature
  introduced in \cite{pavicic-pra-17}.}
\label{figure-3}
\end{figure*}
\begin{figure*}[htp]
\begin{center}
\includegraphics[width=.99\textwidth]{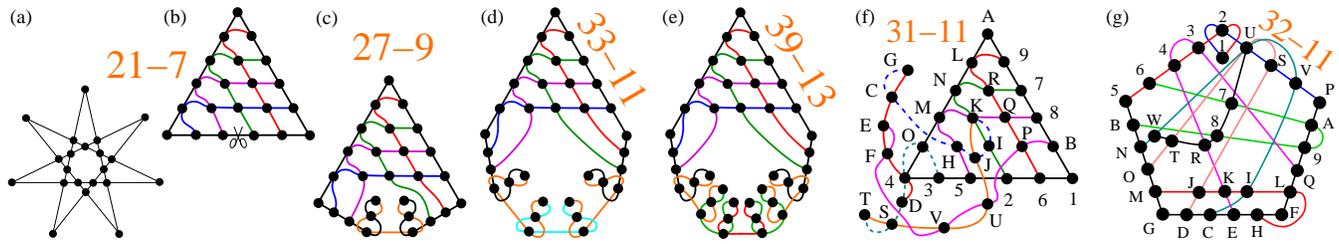}
\end{center}
\caption{6D $\omega$ KS criticals sets; (a) a star representation
  of 21-7 from \cite{lisonek-14}; (b) its isomorphic triangular
  equivalent from \cite{pavicic-pra-17}; (c,d) the only two
  other criticals in the $\omega$ class within the
  $\omega$-coordinatization; pairs of $\delta$-triples
  comes to the triangle at the scissor indicated point;
  (e) 39-13 generated within the $\omega^2$-coordinatization
  from the 591-1123 master set and is contained neither in the
  $\omega$ 216-153 nor in the $\omega^2$ 81-162 class; (f) the
  second smallest from the 81-162 class presented so as to show
  the ``remnants'' of the 21-7 triangle; (g) the third smallest
  set from the class shown in the standard maximal loop
  representation; MMP hypergraph encodings and their
  coordinatizations are given in the Appendices.}
\label{figure-4}
\end{figure*}
\begin{figure*}[htp]
\begin{center}
\includegraphics[width=.79\textwidth]{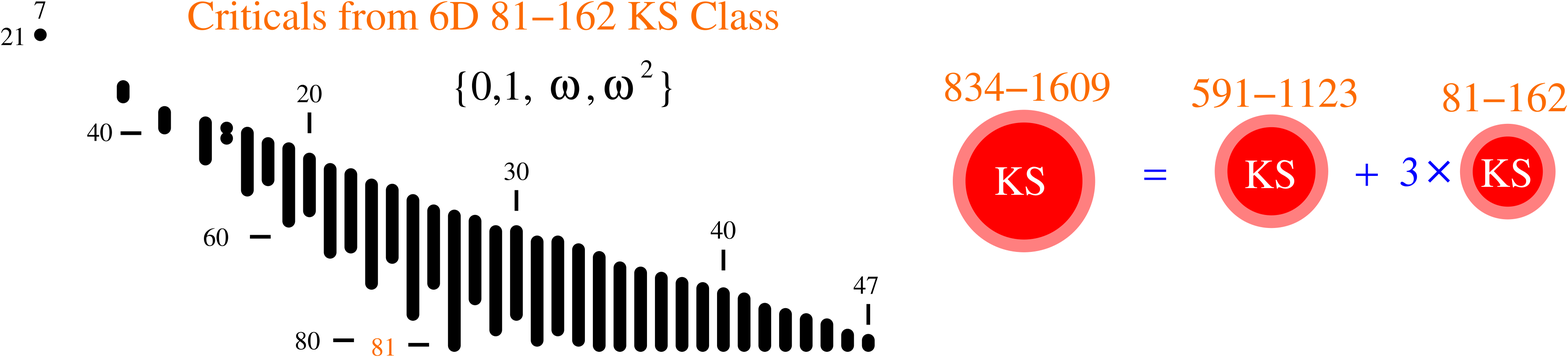}
\end{center}
\caption{Distribution of $2.5\times 10^7$ non-isomorphic KS
  criticals from the 6D $\omega^2$ 834-1609 KS class. See the
  caption of Fig.~\ref{figure-1} for the representation of the
  criticals via dots and lines in the main figure. Vertex units
  are scaled to 1/4 of edge units for a compact presentation.
  The 834-1609 master consists of two submasters---see text.}
\label{figure-5}
\end{figure*}
\end{widetext}

\section{\label{subsec:6dim}Extension to Higher Dimensions}

An automated generation of KS sets of the kind presented in the
previous section can be extended to any dimension. Here we give
an example whose generation escaped all previous attempts via
other methods. It is related to the star-like 6D 21-7 KS set
\cite{lisonek-14} which was implemented by \cite{canas-cabello-14}.
They made use of $\omega^2$-coordinatization (where
$\omega=e^{2\pi i/3}=(-1+i\sqrt{3})/2$) defined by the vector
components from the set $\{0,1,\omega,\omega^2\}$. In
\cite{pavicic-pra-17} it was shown that the set can be given a
triangular hypergraph representation and a simpler
$\omega$-coordinatization based only on the components from the set
$\Omega=\{0,1,\omega\}$. All previous attempts
\cite{lisonek-14,pavicic-pra-17} to find a KS class, or a master set
to which the 21-7 set might belong, failed. In \cite{pavicic-pra-17},
a huge polytope based 6D class was generated, but the 21-7 provably
did not belong to it. The vector generation based on the latter
$\omega$-coordinatization gives an $\omega$-master set 216-153
and its class right away, though. The master set is connected,
i.e., it consists of a single KS set and does not contain any
unconnected non-KS sets as the master sets shown in
Figs.~\ref{figure-1},\ref{figure-2} do. It means that the
$\omega$-coordinatization exhaust all possible assignments from
$\Omega$. Still, the class 216-153, apparently contains only
three critical KS sets, shown in Fig.~\ref{figure-4}(a,b,c),
meaning that after several CPU years of generating criticals
from the master set 216-153 we did not obtain any other
critical---only a myriad of isomorphic replicas of those
three. All three of them have parity proofs.

A natural question arises:\ is there at least the
next critical in the series, i.e., the 39-13 one with an additional
pair of $\delta$-triples, with however small probability of
appearing? (By the very definition of the MMP hypergraphs, 6D ones
can have up to four vertices that share two edges.) The answer is
in the negative, because ascribing of the $\omega$-coordinatization
to its hypergraph fails.

But we are able to vector generate 39-13
from the $\omega^2$-coordinatization---the (d)-heptagon from
Fig.~\ref{figure-4} is then extended to a nonagon with two
additional $\delta$-triples at the bottom as shown in
Fig.~\ref{figure-4}(e).

What geometrical property of the
216-153 master limits it to just three criticals 21-7, 27-9,
and 33-11 and why the generation of the 39-13 critical requires a
bigger $\omega^2$-coordinatization, remain open questions.

The  $\omega^2$-coordinatization generates an 834-1609 master set.
Unlike the $\omega$-master set 216-153, the 834-1609
$\omega^2$-master set consists of four unconnected KS sets
as shown in Fig.~\ref{figure-5}---one 591-1123 and three
mutually isomorphic 81-162 sets, and thus amounts to just two sets.
We have not yet generated the 591-1123 KS criticals since
it requires too many CPU years on supercomputers due to their
numerosity and dimensionality.

\section{\label{sec:disc}Discussion}

In this paper we designed an unprecedentedly fast and
efficient way of generating Kochen-Specker (KS) sets, a kind of
quantum contextual sets, for application and implementation in
quantum communication and computation. The design is based on our
discovery \cite{pm-entropy18} that fairly simple sets of vector
components can generate, with a suitable algorithm, master KS sets
within seconds on any PC. In this paper we provide methods of 
exhaustive generation of smaller critical KS subsets from such master
sets. We call this kind of generation of both master KS sets and
smaller KS sets they contain, the
{\em vector generation of {\rm KS} sets\/}.

We present it in detail for 4D Hilbert space, although
the method is applicable to any dimension. As an example, we
provide a generation of the star/triangle 6D class which the
previous methods failed to provide. We also discuss a 3D case.

With vector generation, we effectively circumvent solving systems
of nonlinear equations to find coordinatizations for generated
KS hypergraphs that we faced in our previous approaches starting
with \cite{pmmm05a}. In the latter approach, we first generated
MMP hypergraphs and then searched for possible coordinatizations
for them. The vector generation approach is its inverse---we first
generate a coordinatization and then search for possible MMP
hypergraphs for them. This reduces finding of KS masters
together with their coordinatizations from CPU years of
calculations on supercomputers to mins on a PC.

At the present stage of our research, we combine two ways of
finding sets of simple vectors for vector generation of KS
sets: random computer searching, and taking over the
coordinatization of master KS sets previously found via polytope
symmetries or Pauli operator structures. In order to find an optimal
approach to a successful vector generation of KS master set we have
to find a balance between desired features of the KS classes we want
to generate from a chosen master set and the required complexity of
the their computer generation on supercomputers via our programs.
For this purpose we extract many characteristic features of the
sets while processing them.

We found that vector components exhaustively generate a
coordinatization of a KS master in the sense of employment of
{\em all\/} possible vectors. This, in most cases, results
in a master set split into several unconnected sets as shown in
Figs.~\ref{figure-1}, \ref{figure-2}, and
\ref{figure-4}. When a master splits into two unequally big
KS submasters as in the 6D example given in
Fig.~\ref{figure-4} this enables us to generate smaller
criticals because the original master 834-1609 and/or the bigger
submaster 591-1123 are computationally unfeasible on a supercomputer.
In 4 dimensions, vector generation of criticals from the 972-1852
KS master set, i.e., the 636-1657 submaster, is perfectly feasible
as shown in Figs.~\ref{figure-2}. Also the classes 300-675
and 148-265 (see the Appendices), even when enlarged
in volume and scope by the vector generation of master sets which
contain them, stay computationally feasible.

We also obtained and relied on a number of generic features of
KS sets.

Complex components of vertices of the 4D KS classes are in the
majority of cases linked to their implementation by means of two
qubits, on the one hand, and to their $\delta$-feature, on the
other, where the latter feature means that pairs of edges share
two vertices, i.e., intersect each other twice. Exceptions are
among others, the smallest KS sets which can acquire both complex
and real coordinatization as, e.g., the 18-9 KS set
\cite{cabell-est-96a}. Real components of vertices mostly lead to
spin-$\frac{3}{2}$ particle implementation, in the sense that
corresponding operators are not qubit-describing
${\mathcal{H}}^2\otimes{\mathcal{H}}^2$ Pauli operators.

The $\delta$-feature stems directly from the $n-2$ hypergraph
condition within the definition of MMP hypergraphs given in
Subsec.~\ref{subsec:mmp}. This implies that edges of 3D MMP
hypergraphs can intersect each other in at most $3-2=1$ vertex.
In addition, in \cite{pmmm05a}, we proved that the size of minimal
loops of adjacent edges is not 2 as in 4 or 6D (exhibiting the
$\delta$-feature as in Fig.~\ref{figure-3}), but 5 (pentagon). These
two features make 3D MMP hypergraphs computationally different from,
e.g., 4D ones. Thus, although we can obtain 3D master hypergraphs as
we pointed out above, we first have to develop new algorithms that
would enable us to generate 3D criticals from them. This is a work in
progress, which is the reason why we have not included any 3D
results in this paper.

As a general rule, also for higher dimensions \cite{pavicic-pra-17},
bigger KS sets do not have parity proofs at all, while the very
small ones with odd number of edges all have them, as shown in
Figs.~\ref{figure-1} and \ref{figure-2}. Among the criticals
of the 148-265 class, which does not have really small sets, not a
single set with a parity proof was generated.

Sets generated from small master sets, like 4D 60-74, 60-105,
73-? and 84-?, and 6D 81-162 exhibit a flat cutoff as shown in
Figs.~\ref{figure-1}(a), \ref{figure-2}(b),
\ref{figure-5}, and \ref{figure-7}, and form a saw-toothed
distribution as shown in  Fig.~\ref{figure-7}(b); this feature
enabled us to fill in the gap between the upper and lower
clusters of the 300-675 class.

All classes generated from sufficiently big master sets, exhibit a
spindle-like distribution of sets with an agglomeration of criticals
below the line that connects the smallest and largest of them as
shown in Figs.~\ref{figure-1}(a), \ref{figure-2}(a), and
\ref{figure-5}.

The 96-96 KS set is found to properly contain all sets from the
60-74 (60-75) class. The big gap between 82 (41) and 211 (127)
vertices (edges) in the 300-675 KS class we obtained in
\cite{megill-pavicic-mipro-17,pavicic-pra-17} is filled in
as shown in Fig.~\ref{figure-1}. The 96-96 KS sets form a subclass
of the 300-675 KS class. Missing KS sets from the 147-265 KS class in
\cite{megill-pavicic-mipro-17,pavicic-pra-17} are obtained.

In 6D the $\omega$-coordinatization of the star/triangle 21-7 KS
set \cite{lisonek-14,pavicic-pra-17} is found to generate a
master set 216-153 which contains only three criticals 21-7, 27-9,
and 33-11 shown in Fig.~\ref{figure-4}(a-d). The
$\omega^2$-coordinatization generates two KS sub-classes 591-1123
and 81-162; the distribution of the latter is shown in
Fig.~\ref{figure-5}. The former is too big to be
generated within the scope of this paper, but we  confirmed that
the 39-13 critical shown in Fig.~\ref{figure-4}(e) belongs
to the former and not to the latter class.

In the end, we would like to recall that just over 50 years have
passed since Simon Kochen and Ernst P. Specker formulated their
KS theorem \cite{koch-speck} and that for 36 years thereupon only
three 4D KS sets were found. The results obtained since then
and partly in this paper prove the power of computer science in
and interdisciplinary approaches to the field.

\section{\label{sec:meth}Methods}

The MMP hypergraph language, and that algorithms and programs
developed for it, are our general method for processing and
categorising KS sets and other properties represented by
hypergraphs. By using exactly one line of text for each MMP
hypergraph, we manipulate a collection of millions of MMPs with
standard text processing tools as well
as distribute pieces of the collection to different CPUs for
massive parallel processing.  Most of our programs work with
arbitrary hypergraphs and can be useful for any project using
hypergraphs for knowledge representation. While we encourage the
use of MMP notation because it is well-defined and convenient, we
can also translate to and from other hypergraph representations.

All our programs are freely available from our repository
\cite{master-repository-17}, and with them a researcher can
reproduce the results reported in this paper. Any of our
programs can process an arbitrary
number of lines each consisting of ASCII characters that represent
KS sets in the form of KS hypergraphs. There are no inherent
limitations in the number of vertices (vectors) or edges in which
vertices are organized (mutually orthogonal $n$-tuples of
vectors determining the
dimension $n$ of the Hilbert space in which vectors reside).

Our programs have many different options available, which are
documented via the {\tt {-}{-}help} for each of them. Their most
important functions are as follows. Program {\textsc{Vecfind}}
generates master sets from the input vector components, like
\{-1,0,1\}, or verifies whether given KS hypergraphs can be
assigned a coordinatization based on such sets. {\textsc{Mmpstrip}}
strips a specified number of edges from a given KS sets or adds them.
{\textsc{Shortd}} reduces collections of KS sets to non-isomorphic
ones, thus eliminating duplicates. {\textsc{Mmpsubgraph}} verifies
whether a given hypergraph is a subgraph of another bigger one.
{\textsc{States01}} filters out KS sets and critical KS sets from
a given collection of sets. {\textsc{Loop}} generates the biggest
loop for a given KS set which enables us to draw its figure by a
program written in {\textsc{Asymptote}} or manually in
{\textsc{Xfig}}. Note that one can easily read the ASCII string KS
hypergraph representation off any figure even when no ASCII
characters are assigned to vertices in the figure. What
characterises a KS hypergraph is its structure, not a specification
of characters or coordinates assigned to vertices. All these
programs are written in C, developed in
\cite{bdm-ndm-mp-1,pmmm05a,pm-ql-l-hql2,pmm-2-10,bdm-ndm-mp-fresl-jmp-10,mfwap-11,mp-nm-pka-mw-11,megill-pavicic-mipro-17},
and extended here. We also used parity-proof algorithms and
programs developed in
\cite{aravind10,waeg-aravind-jpa-11,waeg-aravind-megill-pavicic-11,waeg-aravind-jpa-15}.
Programs allow standard inputs and outputs and can
be piped into each other.

\begin{acknowledgements}

Supported by the Croatian Science Foundation through project
IP-2014-09-7515, the Ministry of Science and Education
of Croatia through the Center of Excellence for Advanced
Materials and Sensing Devices (CEMS) funding, and by
MSE grants Nos. KK.01.1.1.01.0001 and 533-19-15-0022.
Also supported by the Alexander or Humboldt Foundation.
Computational support was provided by the cluster Isabella of
the Zagreb University Computing Centre, by the Croatian National
Grid Infrastructure (CRO-NGI), and by the Center for Advanced
Computing and Modelling (CNRM) for providing computing resources
of the supercomputer Bura at University of Rijeka in Rijeka,
Croatia. The supercomputer Bura and other ICT research
infrastructure were acquired through the project
{\it Development of research infrastructure for laboratories of
the University of Rijeka Campus} which is co-funded by the
European regional development fund. The supports of Emir
Imamagi\'c and Daniel Vr\v ci\'c from Isabella and CRO-NGI and
of Miroslav Pu\v skari\'c from CNRM to the technical work are
gratefully acknowledged. This research was also supported (in part)
by the Fetzer-Franklin Fund of the John E. Fetzer Memorial Trust.

\end{acknowledgements}

\section*{Author contributions statement}

M.P.~wrote the main part of the manuscript, wrote the programs for
supercomputers, generated part of the data on supercomputers
(mentioned in the Acknowledgements), and
prepared all the figures. M.W.~and P.K.A.~wrote parts of the
manuscript and generated a part of data from the programs they
wrote on their computers. N.D.M.~wrote parts of the manuscripts
and a number of programs used in the paper.

\section*{Data availability and Competing interests}

Our programs are freely available at our repository http://goo.gl/xbx8U2.
The datasets generated during the current study are available from the
corresponding author on reasonable request.

The authors declare that there are no competing financial or
non-financial interests in relation to the presented work. 

\appendix

\section{\label{sec:app1}{\{-1,0,1\}-Coordinatization:
    40-32 KS Class}}

Program {\textsc{Vecfind}} with option {\tt{master}} gives the
following encoding of the {\bf 40-32} master hypergraph:
{\tt v40e32} 1234,1256,1378,19A4,23BC,2DE4,FG34,FG56,\break F5HI,FJK6,G5LM,GNO6,DE78,D7OI,D8JL,E7KM,\break E8NH,9ABC,9BOH,9CKL,PQRS,PTUV,WXRY,WZaV$\!$,\break ABJM,ACNI,bXUc,bZdS,eQac,eTdY,NOHI,JKLM.\break \{1=\{0,0,0,1\},2=\{0,0,1,0\},F=\{0,0,1,1\},G=\{0,0,1,-1\},3=\break \{0,1,0,0\},D=\{0,1,0,1\},E=\{0,1,0,-1\},9=\{0,1,1,0\},P=\{0,\break 1,1,1\},W=\{0,1,1,-1\},A=\{0,1,-1,0\},b=\{0,1,-1,1\},e=\{0,\break -1,1,1\},4=\{1,0,0,0\},B=\{1,0,0,1\},C=\{1,0,0,-1\},7=\{1,0,\break 1,0\},X=\{1,0,1,1\},Q=\{1,0,1,-1\},8=\{1,0,-1,0\},T=\{1,0,-1,\break 1\},Z=\{-1,0,1,1\},5=\{1,1,0,0\},a=\{1,1,0,1\},U=\{1,1,0,-1\},\break d=\{1,1,1,0\},N=\{1,1,1,1\},J=\{1,1,1,-1\},R=\{1,1,-1,0\},\break K=\{1,1,-1,1\},O=\{1,1,-1,-1\},6=\{1,-1,0,0\},S=\{1,-1,0,1\},\break Y=\{-1,1,0,1\},V=\{1,-1,1,0\},L=\{1,-1,1,1\},H=\{1,-1,1,-1\},\break c=\{-1,1,1,0\},I=\{1,-1,-1,1\},M=\{-1,1,1,1\}\}.

\begin{figure}[htp]
\begin{center}
\includegraphics[width=.47\textwidth]{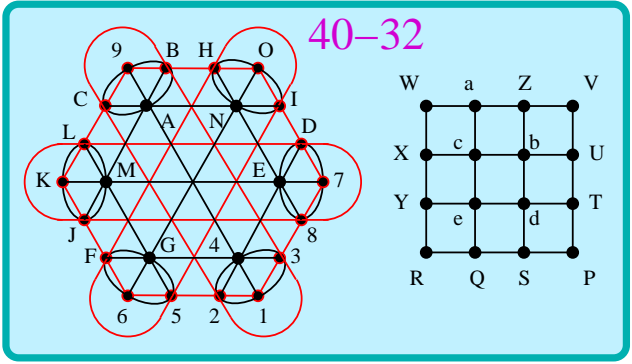}
\end{center}
\caption{Graphical representation of the 40-32 master set with
  \{-1,0,1\}-coordinatization. It consists of two disconnected
  hypergraphs: the left one is isomorphic to Peres' 24-24 KS
  hypergraph (\cite{peres,pmm-2-10,pavicic-pra-17}) and the right
  one is a non-KS hypergraph which is generated to exhaust
  \{-1,0,1\}-coordinatization. The smallest 4D KS set 18-9, first
  found in \cite{cabell-est-96a}, is indicated in red. For the
  other five critical KS sets see \cite{pavicic-pra-17}.}
\label{figure-6}
\end{figure}

Its graphical representation is given in Fig.~\ref{figure-6}

Program {\textsc{Mmpstrip}} generates 3,712 smaller KS sets
which together with their 40-32 master set form the 40-32 KS class.
Of those, 1,233 connected ones (up to 24 vertices and 24 edges) are
isomorphic with the ones previously obtained from Peres' 24-24 KS
set in \cite{pmm-2-10}.

\section{\label{sec:app2}Polytope Generation}

\subsection{\label{subsec:app2-1}600-Cell Generates the
  60-74 KS Master Set}

The 600-cell has 25 different 24-cells inscribed in it, in the way
that a regular dodecahedron has five cubes inscribed in it. The
inscribed 24-cells can be seen most easily in the basis table of
the 600-cell, where each 24-cell can be seen as one of the blocks
of 12 vectors (each representing an antipodal pair of vertices of
this 24-cell). The vertices of the 600-cell can be partitioned into
those of five disjoint 24-cells in ten different ways, with the rows
and columns of Table 2 of \cite{waeg-aravind-megill-pavicic-11}
showing the different partitions. It is important to note that
although the 600-cell has many 24-cells in it, no 24-cell is
accompanied by its dual. Thus, the 24-24 system described above is
not a subset of the 60-75 master set.

\subsection{\label{subsec:app2-2}120-Cell Generates the
  300-675 KS Mater Set}

The 120-cell \{5,3,3\} has 120 dodecahedra for its outer boundary,
with three around each edge of the polytope. Alogether it has
120 pentagons, 1200 edges, and 600 vertices. These vertices, lying on
a 3-sphere in 4D, come in antipodal pairs, and so give rise to just
300 distinct vectors or quantum states. These 300 vectors form 675
bases of four mutually orthogonal states, from which we obtain the
300-675 KS master set according to the prescription given in
\cite{waeg-aravind-fp-14}.

\subsection{\label{subsec:app2-3}Polytope Generation of
  the 300-675 and 96-96 KS Classes and Their Features}

In \cite{waeg-aravind-fp-14} we obtained a 300-675 KS master
set from the 120-cell polytope. However, the set was too big
for our parity proof program, so that we made use
of a 96-96 subset to generate critical (smallest) KS sets
\cite[Table 5]{waeg-aravind-fp-14}. It is not too big for
our {\textsc{Mmpstrip}}, {\textsc{States01}} and
{\textsc{Mmpsubgraph}} programs
\cite{pmmm05a-corr,pmm-2-10}, though, and we obtained all
the results in this section by making use of these three
programs. The MMP hypergraph string of the master set 300-675
itself is however too long to be reproduced here and the
reader can retrieve it from our repository in
\cite{master-repository-17}.

Subsets of the master set 300-675 form the 300-675 KS class and
in Fig.~\ref{figure-1} we give all the critical KS sets we
obtained from this class. 

The very construction of the 300-675 master set (presented in
\cite{waeg-aravind-fp-14}) involved the 60-75 master set and
therefore it constructively proved that the 300-675 KS class
properly contains the 60-74 class. Namely, by stripping one
edge at the time from the 60-75 master set we obtain seventy
five 60-74 KS subsets. They turn out to be isomorphic to each
other and actually merge into a single 60-74 set. Thus, both
sets can be taken as master sets for smaller KS sets which
form a class of their subsets and in the literature we called
them 60-75 and 60-74 classes, respectively. They are identical,
short of the very 60-75 set from the 60-75 class. This is a
general feature of any master set in which all edges are
symmetrically identical. 

In \cite{waeg-aravind-fp-14} we called the 60-75 master set
(first discovered in \cite{aravind10} and further elaborated in
\cite{waeg-aravind-megill-pavicic-11,mfwap-11}), the 600-cell.
In that paper we mostly considered the 96-96 set (``one of the
reduced sets,'' actually the last one from Table 4 of the paper,
i.e., $36_248_512_6-96_4$). It turns out that the 96-96 set
properly contains the 60-75 set and therefore also the 60-74 set
as well as all the sets from the 74 class. To see this, let us
have a look at Table 1 of \cite{waeg-aravind-fp-14}. In the table,
the 60 rays of the 60-75 are labelled E$'$ (they form a subset of
the 300-675). We can construct the 96-96 so as to cancel
12-ray-blocks and from Table 4 we then see that E$'$ remains intact.

\begin{figure*}[htp]
\begin{center}
\includegraphics[width=.9\textwidth]{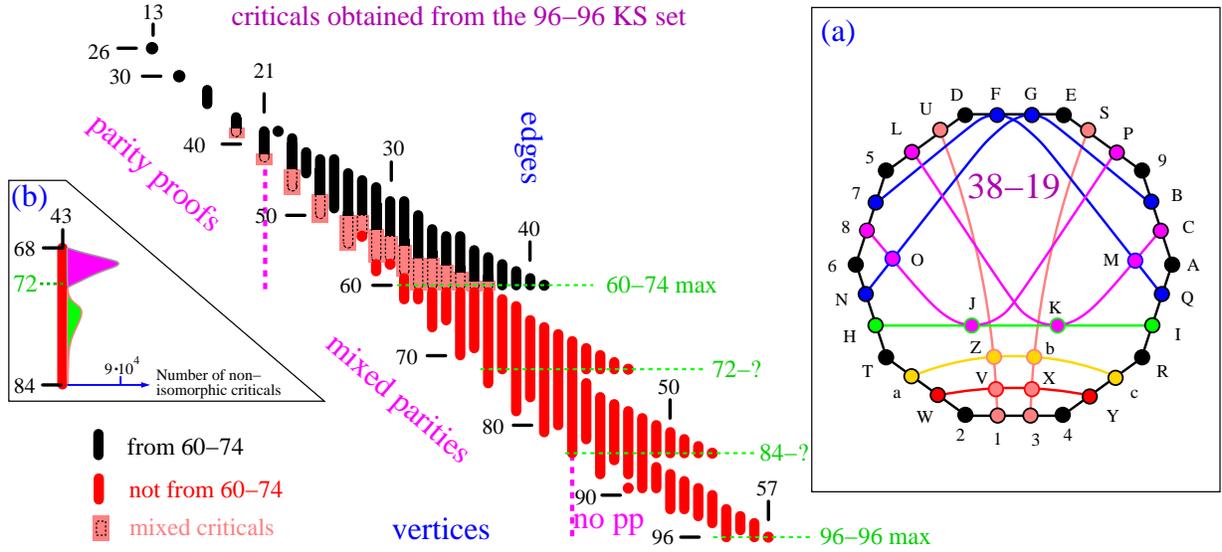}
\end{center}
\caption{Distribution of $5\times 10^9$ KS criticals obtained from
  the 96-96 KS subset of the 300-675 KS master set
  \cite{waeg-aravind-fp-14} via {\textsc{Mmpstrip}} and
  {\textsc{States01}} programs.
  For an explanation of differently colored criticals see text.
  Vertex units are scaled to 1/2 of edge units to give a more
  compact graphical presentation. ``Mixed criticals'' mean that
  only a portion of KS criticals with the shown vertices and edges
  are from the 60-74 class. Inset (a): the smallest critical
  hypergraph which is not a subgraph of the 60-74 master
  KS set. Inset (b): the distribution of all criticals
  with 43 edges. ``no pp'' means ``no parity proofs.''}
\label{figure-7}
\end{figure*}

Another way to prove that the 96-96 subset contains the 60-75
master set and all KS criticals from the 60-74 class in addition
to criticals not contained in the 60-74 class is to make use of
our universal programs {\textsc{Mmpstrip}} and {\textsc{States01}}.
In this approach we obtain a number of results which are presented
below and partly displayed in Fig.~\ref{figure-7}.

The black criticals are from the 60-74 class. Squared pink
rectangles with dashed lines mean that some of the criticals are
from the 60-74 while the others are not. Criticals in red do not
belong to the 60-74 class. We obtained all the criticals via the
universal {\textsc{Mmpstrip}}, {\textsc{States01}}, and
{\textsc{Mmpsubgraph}} programs but we discerned 60-74 from
non-60-74 criticals with an odd number of edges mostly via our
parity proof program which is much faster than
{\textsc{Mmpsubgraph}}. Criticals with an even number of edges
which cannot be discerned by means of parity proof algorithms
we discerned via {\textsc{States01}}. Criticals from the 60-74
cannot have more than 60 vertices which we visualized by means of
the ``60-74 max'' dashed line. The 300-675 class contains the
60-74 class but it does not exhibit a cutoff shape.
Its MMP hypergraphs contain edges with fewer and fewer vertices
toward the opposite ends of their distribution. Also they
exhibit agglomeration of criticals below the line which connects
the smallest and largest hypergraphs. As we can see in
Fig.~\ref{figure-9}(b), the 148-265 class possesses the same
features.

The above cutoff shape of the 60-74 class appears to be
characteristic of any subclass of a bigger class, which we
tentatively call a {\em complete class\/} generated from a
{\em complete master set\/}. Under {\em complete sets\/} we
understand the ones obtained from an exhaustive usage of
vector components to build their coordinatization. Then
their distribution exhibits a spindle-shaped agglomeration
elaborated on in Subsec.~\ref{subsec:vgen}. E.g., in
Fig.~\ref{figure-7}, from the distribution of the 96-96 criticals,
in particular from its saw-toothed shape, we can, apart from the
subclass 60-74, recognize two new subclasses 72-? and 84-? where
``?'' stands for the edges of the sub-master sets that we do not
know as of yet. Also, through the cutoff at the 96 vertex level,
we can see that the 96-96 class itself is a subclass of a bigger
complete set, e.g., the 300-675 one. Our claim that the saw-toothed
shapes are genuine effects and not consequences of insufficiently
intensive generation of sets are supported by, first, the fact
that among the last 15\%\ of generated sets there were no new
types belonging to the teeth area (only two from bottom left
area), and, second, by two clearly pronounced sub-distributions
of vertices related to the same edges and the base of each tooth.
In Fig.~\ref{figure-7}, an example of such a double sub-distribution
is shown in the Inset (b) for the edge 43. There is only one KS
{\bf 73-43} critical (enneadecagon in the lower runs of the program
{\textsc{Loop}}): {6587,7wxr,rq1n,nlmk,koec,cb4P,PKGC,C9BA,AIEU,\break U\#s3,3$*$"g,gaYW,WVOT,TRSQ,Q/\$z,zy!f,f(\%d,d\&h$'$,\break $'$u26,,,1234,DEFG,HIJK,LMNO,JFB8,XYSN,ZaRM,\break de73,fgc6,hi54,jHD9,iZXV,opL2,stpn,uvie,$"$!rb,\#vmb,\break \$\%t4,)\$l6,$*$(o5,-/q5,\#zxh,-)db,/\&sc}.

Another example of the distribution shown in the Inset (b) of
Fig.~\ref{figure-7}, is of sets with 37 edges with an order of
magnitude bigger sub-distribution of vertices than for the above
43 edges.

MMP hypergraph representations of the smallest KS criticals
with an odd number of edges that do not belong to the 60-74
class, obtained from the 96-96 sets with parity proof
programs, in particular two 38-19, 42-21, and 48-25,
are given in \cite[Fig.~8]{pavicic-pra-17}. In Fig.~\ref{figure-7},
in the inset (a), the hypergraph figure of 38-19 is given.
Its code reads: {\bf 38-19}: {2134,4cYR,RQIA,ABC9,9SPE,\break
EFGD,DUL5,5786,6NHT,TaW2,,,HIJK,LMKC,NOGB,\break POJ8,QMF7,VWXY,Zabc,bXS3,ZVU1.}

\begin{figure*}[htp]
\begin{center}
  \includegraphics[width=.99\textwidth]{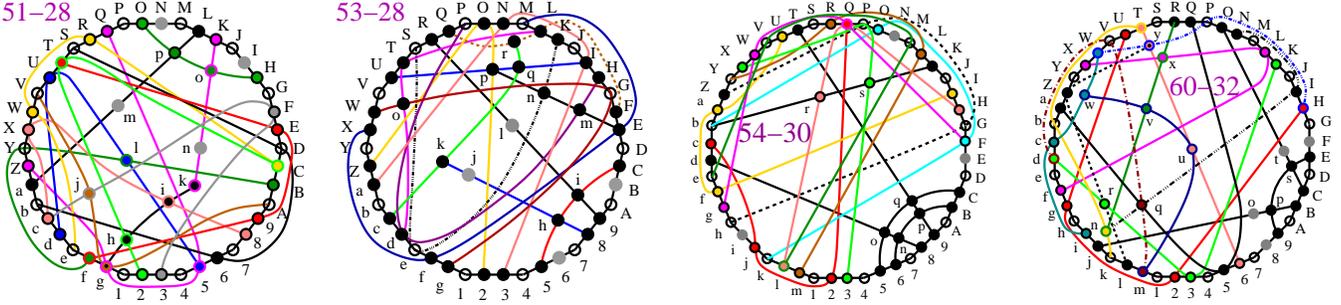}
\end{center}
\caption{Four small KS criticals with an even number of edges.
  In contrast to small criticals with an odd number of edges
  they all have several vertices that belong to only one edge
  (denoted as grey dots).}
\label{figure-8}
\end{figure*}

In Fig.~\ref{figure-8} we give some MMP hypergraph
representations of the smallest KS criticals with an even
number of edges, which cannot be obtained by means of
parity proof algorithms.

They lack the symmetry many of KS criticals with an odd number
of edges exhibit (Cf.~\cite[Figs.~3 and 8]{pavicic-pra-17}).
Their MMP hypergraphs run as follows.

\begin{widetext}

\parindent=0pt
{\bf 51-28}: {1234,4567,789A,ABCD,DEFG,GHIJ,JKLM,MNOP,PQRS,STUV,VWXY,YZab,bcde,efg1

\parindent=30pt
,,,2hTC,3Fjc,5lUd,5QZg,6bDS,8iXc,9fET,AgjV,BlYf,CRWe,HopO,Konk,Lpma,ghik.}

\parindent=0pt
{\bf 53-28}: {,,,aNI3,XMfE,5hiC,8hjk,9ilR,Emnp,gGoW,SenK,PdmH,UpqI,ckqr,QrLF,doTL,ZOp2.}

{\bf 54-30}: {,,,2Qjc,3sPX,5oqC,6npB,8nod,9pqT,FQWg,FkOb,GhPZ,HQrl,IfaU,Ksrb,LmRY,MlVe.}

{\bf 60-32}: {,,,2HUg,3Krd,5Qqa,6uyT,9psE,Bpoj,CstO,HPyW,INqn,LxXf,Rxvn,Tbnk,Vdqm,Wwci,Zrly,wvum.}

\end{widetext}

\parindent=20pt

The main loops are obvious from the figures (alphabetical
order), and except for the first set we omit them. The
loops are 14-gons for the first two and 16-gons for the
last two sets. Some 58-32 and 59-32 have 17-gons as maximal
loops but these are actually small as compared to many
KS criticals from the upper part of the class shown in
Fig.~\ref{figure-1}, whose maximal loops are 55-gons and bigger.

In \cite{pavicic-pra-17,megill-pavicic-mipro-17} a uniform
random generation of criticals with {\textsc{Mmpstrip}} and
{\textsc{States01}} generated a cluster of sets with edges
between 127 and 188 and vertices between 211 and 283.
The programs did not generate a single critical in between
this cluster and the 96-96 cluster. When we tried to find
out the cause of the gap we first realised
that none of the criticals from the upper cluster had a parity
proof, so, we could not obtain them with a parity proof program.
Then, we realised that a probability of finding criticals between
the aforementioned criticals via a uniform random procedure in
the 300-675 class is extremely low (unlike in any other class in
any dimension \cite{pavicic-pra-17}) and that a random generation
of each of them might take up to 3 months. However, following the
above 96-96 saw-toothed feature, we conjectured that an analogous
feature repeats throughout the 300-675 class up to the upper cluster.

We verified the conjecture in the following manner. We first let
100 jobs run in parallel for 3 months and obtained a number of
criticals in between the former upper and lower clusters. Then we
traced the subsets they were generated from. When a considerable
number of criticals pointed to particular subsets, we ran up to 200
{\textsc{States01}} jobs on these subsets in parallel. This proved
fruitful and we filled in the gap, denoted by question marks in
Fig.~9 \cite{pavicic-pra-17} and Fig.~2 of
\cite{megill-pavicic-mipro-17}, and obtained evenly distributed
critical KS set shown in Fig.~\ref{figure-1}.

We found no parity proof among criticals bigger than 82-41,
so the upper part of the 300-675 KS class is invisible to
the parity proof algorithms.

The biggest maximal loop we found is a 57-gon for a 283-188 critical.

The 300-675 KS class has a real coordinatization, i.e.,
vectors corresponding to vertices of their MMP hypergraphs
are from $\mathbb{R}^{4}$. The consequence is
the structural simplicity of the hypergraphs. In particular,
they all lack the $\delta$-feature of two edges sharing
two vertices characteristic of the hypergraphs from the
60-105 class that has the complex coordinatization.

\subsection{\label{subsec:app2-4}Witting Polytope
  Generates the 148-265 KS Class}

Complex coordinatization does not automatically
furnish the hypergraphs with intricate features, as the 148-265
class shows \cite{megill-pavicic-mipro-17,pavicic-pra-17}.
The 148-265 KS class sits in complex 4D Hilbert space, and
the two classes 300-765 and 148-265 are completely disjoint,
making the latter class particularly
interesting for designing experiments with qubits residing
in complex Hilbert space. The class is obtained from a polytope
with coordinatization in $\mathbb{C}^{4}$, called the
{\em Witting polytope}.

The well-known Penrose dodecahedron 40-40 KS set in 4D complex
Hilbert space is related to the Witting Polytope, through the
Majorana representation of spin states to the regular
dodecahedron in 3-dimensional Euclidean space, and through a
particular projection, to the root structure of the Lie algebra
E8 in the 8-dimensional Euclidean space
\cite{waeg-aravind-pla-17}. Its coordinatization can be obtained
from the set $\{0,\pm 1,\pm\omega,\pm\omega^2\}$, where
$\omega=e^{2\pi i/3}$.

Penrose's 40-40 set is not critical, and we exhaustively
generated all KS subsets it contains. The outcome turns out
to be of importance for the definition of critical sets
since all criticals contain all 40 vertices of the master
set and are obtained by removing only edges.  This shows that
the vertex criticality criterion ``A KS set is critical if by 
deleting any of its vertex it turns into a non-KS set'' which
can be found in the literature (for example
\cite{ruuge12,zimba-penrose}) is too restrictive.

On the other hand, it is important because it is an extreme
example of a flat cutoff feature with no vertical (vertex)
distribution of hypergraphs.

None of the 40-40 subsets has a parity proof, so, we made use of
the hypergraph program {\textsc{States01}} to generate them.
The 40-40 master set from \cite[Table 3]{waeg-aravind-pla-17} is
represented by the following MMP hypergraph
{1BLV,bDO7,b4UJ,bGAR,1468,BDKF,BGEI,LUNP,\break LOQS,1A35,D3MZ,4EMc,UK2e,GQ2a,ANCX,O6CY,\break 6KdT,ENd9,Q3dH,M2CW,bVdW,BJCH,UI3Y,GP6Z,\break LRMT,1729,4FQX,EeO5,NaD8,ASKc,VaYc,VXZe,\break J5aT,JS9Z,I7XT,I8SW,P5FW,P7cH,R8eH,RF9Y.}

Via exhaustive stripping by means of the program {\textsc{Mmpstrip}}
and filtering through the program {\textsc{States01}}, we obtain all
subgraphs and all critical subgraphs among them, as shown in
Table \ref{table2}.

\begin{widetext}
  \begin{table}[th!]
    \caption{\label{table2}{\hbox to 0.85\textwidth{\quad All subgraphs,
          and all critical subgraphs among them, generated from the
          Penrose 40-40 dodecahedron hypergraph.}}}
\setlength{\tabcolsep}{2.3pt}
\begin{tabular}{|c | c c c c c c c c c c c c c c c c c c || c c c|} 
\hline
  40-40&\multicolumn{18}{|c|}{all dodecahedron subgraphs (all have 40 vertices)}&\multicolumn{3}{|c|}{criticals}\\
\hline
$edges$& 40 & 39 & 38 & 37 & 36 & 35 & 34 & 33 & 32 & 31 & 30 & 29 & 28 & 27 & 26 & 25 & 24 & 23 & 25 & 24 & 23 \\
\hline
No.~of sets &  1 & 1 & 2 & 5 & 15 & 47 & 160 & 553 & 1870 & 5822 & 16208 & 39593 & 82944 & 144315 & 193818 & 164536 & 24948 & 56 & 60752 & 24265 & 56\\
\hline
\end{tabular}
\vbox to 20pt{\vfill}
\end{table}
\end{widetext}

\begin{figure*}[htp]
  \begin{center}
\includegraphics[width=.8\textwidth]{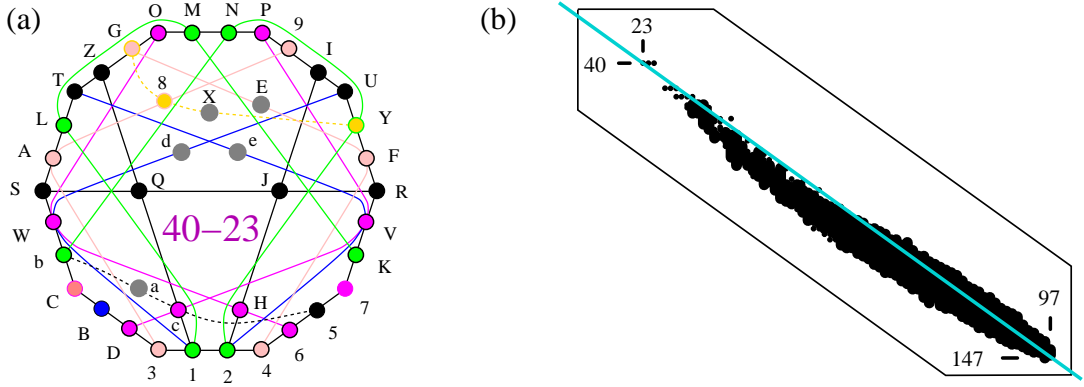}
\end{center}
\caption{(a) 40-23 critical; ASCII symbols of vertices are taken
  directly from the computer generated MMP hypergraph. Grey
  dots depict vertices that belong to just one edge; (b) Distribution
  of 250140 KS criticals from the 148-265 class which shows the
  same kind of agglomeration of criticals below the line which
  connects the smallest and largest of them as in the 300-675
  class in Fig.~\ref{figure-1}. Also, again as in the 300-675 class,
  the number of vertices per edge decreases toward the opposite
  ends of the distribution.}
\label{figure-9}
\end{figure*}

In Fig.~\ref{figure-9} we give a figure of one of the 56 smallest
40-23 KS criticals. Its MMP representation runs as follows (maximal loop
symbols can be read off from Fig.~\ref{figure-9}):
{,,, 89A3,EFG4,HIJ2,KLM1,QRSJ,WOH6,XYZ8,abc5,\break cVPD,cZQ1,bYN2, dWU1,eVT2.} All 40-40 criticals have decagons as their
maximal loops.

The 148-265 KS master set was in \cite{master-repository-17}
constructed with coordinates from
$\{0,\pm i,\pm 1,\pm\omega,\pm\omega^2,\pm i\omega^{1/\sqrt{3}},\break \pm i\omega^{2/\sqrt{3}}\}$ \cite[Table 2]{waeg-aravind-pla-17}.
However, in \cite{pavicic-pra-17}, we showed that the set
$\{0,\pm 1,\pm\omega,\pm\omega^2\}$ suffices.

As shown in Fig.~\ref{figure-9}(b), the criticals we obtained in
the 148-256 class and which do not belong to the 40-40 ones have
vertices in the span from 49 to 147 and edges from 27 to 97 and
exhibit spindle-shaped agglomeration similar to the 300-675 class.
The sets with 49 vertices might belong to a subclass similar
to the 40-40 one, but we have not explored them further.

The maximal loops of the criticals go up to 36-gons, and
their hypergraphs are therefore mostly too big to be efficiently
visualized. One of the smallest non-40-40 ones, a 49-27 critical,
has been presented in \cite[Fig.~10]{pavicic-pra-17} and
\cite[Fig.~2]{megill-pavicic-mipro-17}.

No set we generated from either the 300-675 class or the 148-265 class
contains edges that share more than one vertex.
Some sets from other classes do contain such edges, which
we call the $\delta$-feature, and it proves to be characteristic of the
60-105 class in particular.

\section{\label{sec:app3}6D MMP Hypergraph Encoding}

6D MMP hypergraph encoding of the critical KS set {\bf 21-7} with
$\omega$-coordinatization is given in \cite{pavicic-pra07}
and its $\omega^2$-coordinatization in \cite{lisonek-14}.

MMP encoding of the next two KS criticals (Fig.~\ref{figure-4}(c,d))
are:

{\bf 27-9} {123456,1789AB,CDEFGH,CIJKAL,DMI9NO,\break MF7PKQ,E234GH,R5J8OQ,RPNLB6.\{1=\{0,0,0,0,0,1\},\break 2=\{1,$\omega$,1,0,0,0\},3=\{$\omega$,1,1,0,0,0\},4=\{1,1,$\omega$,0,0,0\},5=\{0,0,\break 0,0,1,0\},6=\{0,0,0,1,0,0\},7=\{0,$\omega$,$\omega$,1,1,0\},8=\{1,1,$\omega$,$\omega$,0,\break 0\},9=\{$\omega$,0,1,$\omega$,1,0\},A=\{$\omega$,1,0,1,$\omega$,0\},B=\{1,$\omega$,1,0,$\omega$,0\},C\break =\{0,0,1,0,0,0\},D=\{0,1,0,0,0,0\},E=\{0,0,0,1,1,$\omega$\},F=\{1,0,\break 0,0,0,0\},G=\{0,0,0,1,$\omega$,1\},H=\{0,0,0,$\omega$,1,1\},I=\{1,0,0,$\omega$,$\omega$,\break 1\},J=\{1,$\omega$,0,1,0,$\omega$\},K=\{0,1,0,$\omega$,1,$\omega$\},L=\{$\omega$,$\omega$,0,0,1,1\},M\break =\{0,0,1,1,$\omega$,$\omega$\},N=\{1,0,$\omega$,0,1,$\omega$\},O=\{$\omega$,0,$\omega$,1,0,1\},P=\{0,\break 1,$\omega$,0,$\omega$,1\},Q=\{0,$\omega$,1,$\omega$,0,1\},R=\{$\omega$,1,1,0,0,$\omega$\}\}}

{\bf 33-11} {123456,789ABC,DEFGHC,IJKLHB,\qquad\break MNOLGA,PQRMNO,STUKF9,VWUJE8,XWTID7,\break PQR456,XVS123.\{1=\{1,0,0,0,1,$\omega$\},2=\{1,0,0,0,$\omega$,1\},3=\break \{$\omega$,0,0,0,1,1\},4=\{0,1,$\omega$,1,0,0\},5=\{0,1,1,$\omega$,0,0\},6=\{0,$\omega$,1,\break 1,0,0\},7=\{0,0,1,1,$\omega$,$\omega$\},8=\{0,1,0,$\omega$,1,$\omega$\},9=\{0,1,$\omega$,0,$\omega$,1\},\break A=\{1,0,0,0,0,0\},B=\{0,$\omega$,1,$\omega$,0,1\},C=\{0,$\omega$,$\omega$,1,1,0\},D=\break \{$\omega$,0,1,$\omega$,1,0\},E=\{$\omega$,1,0,1,$\omega$,0\},F=\{1,$\omega$,1,0,$\omega$,0\},G=\{0,0,\break 0,0,0,1\},H=\{1,1,$\omega$,$\omega$,0,0\},I=\{$\omega$,0,$\omega$,1,0,1\},J=\{1,$\omega$,0,1,0,\break $\omega$\},K=\{$\omega$,1,1,0,0,$\omega$\},L=\{0,0,0,0,1,0\},M=\{0,1,$\omega$,$\omega$,0,0\},\break N=\{0,$\omega$,1,$\omega$,0,0\},O=\{0,$\omega$,$\omega$,1,0,0\},P=\{1,0,0,0,$\omega$,$\omega$\},Q=\break \{$\omega$,0,0,0,1,$\omega$\},R=\{$\omega$,0,0,0,$\omega$,1\},S=\{0,0,0,1,0,0\},T=\{1,0,\break $\omega$,0,1,$\omega$\},U=\{$\omega$,$\omega$,0,0,1,1\},V=\{0,0,1,0,0,0\},W=\{1,0,0,$\omega$,\break $\omega$,1\},X=\{0,1,0,0,0,0\}\}}

Of MMP encoding with $\omega^2$-coordinatization we just give

{\bf 31-11} {123456,1789AB,CDEFGH,CIJ96K,L2M4NO,\break L2EPQH,R7MFGO,R7SPQB,2T8JNU,7TSV5K,\break DI3VAU.\{1=\{$\omega$,1,1,$\omega^2$,$\omega$,1\},C=\{$\omega$,1,1,$\omega$,1,$\omega^2$\},L=\{$\omega$,1,\break 1,$\omega$,$\omega^2$,1\},R=\{1,$\omega^2$,$\omega^2$,1,1,1\},2=\{$\omega$,1,$\omega^2$,1,1,$\omega$\},7=\{$\omega^2$,\break $\omega$,1,$\omega$,1,1\},D=\{$\omega$,1,$\omega^2$,1,$\omega$,1\},I=\{1,$\omega^2$,1,$\omega$,$\omega$,1\},T=\{1,\break $\omega^2$,1,1,1,$\omega^2$\},3=\{$\omega^2$,1,$\omega$,$\omega$,1,1\},M=\{$\omega^2$,1,$\omega$,1,$\omega$,1\},S=\break \{$\omega^2$,1,$\omega$,1,1,$\omega$\},E=\{1,$\omega$,$\omega^2$,$\omega$,1,1\},8=\{1,$\omega$,$\omega^2$,1,$\omega$,1\},J=\break \{$\omega^2$,1,1,1,$\omega^2$,1\},V=\{1,$\omega$,$\omega$,1,$\omega^2$,1\},9=\{$\omega$,$\omega^2$,$\omega$,1,1,1\},P\break =\{1,$\omega$,1,1,$\omega^2$,$\omega$\},F=\{1,$\omega$,1,1,$\omega$,$\omega^2$\},A=\{1,1,1,1,1,$\omega$\},4=\break \{1,$\omega$,1,1,1,1\},5=\{1,1,1,$\omega$,$\omega^2$,$\omega$\},Q=\{1,1,1,$\omega^2$,1,$\omega^2$\},N=\break \{1,1,1,$\omega^2$,$\omega$,$\omega$\},G=\{1,1,1,$\omega^2$,$\omega^2$,1\},6=\{1,1,$\omega$,1,$\omega$,$\omega^2$\},O\break =\{1,1,$\omega$,$\omega$,1,$\omega^2$\},H=\{$\omega^2$,$\omega^2$,1,1,1,1\},B=\{1,1,$\omega$,$\omega$,$\omega^2$,1\},\break K=\{1,1,$\omega^2$,$\omega^2$,1,1\},U=\{$\omega$,$\omega$,1,$\omega^2$,1,1\}\}}

\bibliography{/3rd-disk/ql/m-p}

\end{document}